\documentclass[preprint,showpacs,floatfix,superscriptaddress,jap]{revtex4}
\usepackage{graphicx}% Include figure files
\usepackage{dcolumn}% Align table columns on decimal point
\usepackage{bm}% bold math
\def\nbOne{\ \!\hbox{{\rm 1$\hskip-2.7pt$l}}}

\usepackage{amsmath} % \pmb bold special characters
\usepackage[latin1]{inputenc}
\usepackage[T1]{fontenc}
\usepackage{ascii}
\def\bsigma{{\pmb{\sigma}}}

\begin{document}
\title[Spin filtering device]{Mach-Zehnder Interferometric device for spin filtering in a GaAs/AlGaAs electron gas}

\author{Benjamin Santos}
\address{Departamento de F\'\i sica, Universidad Central de Venezuela, Caracas, Venezuela}
\address{Centro de F\'\i sica, Instituto Venezolano de Investigaciones Cient\'ificas, Apartado 21874, Caracas 1020-A, Venezuela}
\author{Ernesto Medina}
\address{Departamento de F\'\i sica, Universidad Central de Venezuela, Caracas, Venezuela}
\address{Centro de F\'\i sica, Instituto Venezolano de Investigaciones Cient\'ificas, Apartado 21874, Caracas 1020-A, Venezuela}
\address{Statistical Physics Group, P2M, Institut Jean Lamour, Nancy Universit\'e, BP70239, F- 54506 Vand\oe uvre les Nancy, France}
\author{Alexander L\'opez}
\address{Centro de F\'\i sica, Instituto Venezolano de Investigaciones Cient\'ificas, Apartado 21874, Caracas 1020-A, Venezuela}
\author{Bertrand Berche}
\address{Centro de F\'\i sica, Instituto Venezolano de Investigaciones Cient\'ificas, Apartado 21874, Caracas 1020-A, Venezuela}
\address{Statistical Physics Group, P2M, Institut Jean Lamour, Nancy Universit\'e, BP70239, F- 54506 Vand\oe uvre les Nancy, France}

\begin{abstract}
A spin filtering device using quantum spin interference is theoretically proposed in a GaAs/AlGaAs electron gas that has both Rashba and Dresselhaus spin-orbit couplings. The device achieves polarized electron currents by separating spin up and spin down components without a magnetic field gradient. We find two broad spin filtering regimes, one where the interferometer has symmetrical arms, where a small magnetic flux is needed to achieve spin separation, and the other with asymmetric arms where the change in path length renders an extra phase emulating the effects of a magnetic field. We identify operating points for the device where optimal electron polarization is achieved within value ranges found in a 2D electron gas. Both device setups apply for arbitrary incoming electron polarization and operate at broad energy ranges within the incoming electron band.

\end{abstract}

\pacs{72.25.-b,85.75.-d,03.65.Vf}

\maketitle

\section{Introduction}
The Rashba and Dresselhaus spin-orbit (SO) interactions arise in materials which lack either structural or bulk inversion symmetry, respectively\cite{Rashba,Dresselhaus,winkler}. These two kinds of interactions have recently been given a great deal of attention due to their potential role in the generation and manipulation of spin polarized currents, spin filters\cite{Nitta,Ionicioiu,Hatano,SHChen}, spin accumulation\cite{SarmaReview}, and spin optics\cite{BalseiroUsaj}. 

Recent proposals have been made for the construction of perfect spin filters based on active Rashba spin orbit  media\cite{Hatano}, ballistic spin interferometers\cite{Koga} and the analysis of the persistent spin helix\cite{SHChen,Bernevig2}. Further applications based on the interference concept include quantum logic gates\cite{ZulickeAlone}, bit controlled Stern-Gerlach devices\cite{Ionicioiu} and tunable entanglement\cite{SignalZulicke}. Here we readdress the problem of spin filtering by interferometry explicitly implementing the original concept in reference \cite{Ionicioiu,Hatano}, proposing a quasi two dimensional device, to test the spin filtering concept through an electronic Mach Zehnder interferometer (MZI) within Rashba and Dresselhaus media\cite{Lopez,Berche1}. Generally, spin injection is achieved by drawing currents from oriented ferromagnets, where the majority carriers have a preferred spin orientation. In order to change spin orientation it is then necessary to use strong magnetic fields and depend on spin relaxation times that limit manipulation speeds. The currents drawn by these methods still contain a large fraction of the undesired spin orientation depending on the magnetic field and the density of states. For this reason it is very desirable to achieve a larger fraction of oriented spin albeit not a pure spin current. Conceptually, the proposed device would be able to separate spin-orientation, as in a Stern-Gerlach device, but without magnetic field gradients, and both in the absence (asymmetric interferometer) and presence of a weak magnetic field (symmetric interferometer) coupled to the spin-orbit interaction\cite{Berche1}.

The aim of this work is to generate spin polarized currents making use of a combination of weak magnetic fields (where the Zeeman term contribution is small), the intrinsic Dresselhaus spin-orbit interaction and the tunable structure inversion asymmetry (SIA) Rashba type interaction. The interference of electron precession and phase changes in the wave function render sufficient degrees of control to separate spin components into distinct output channels in a variety of experimental parameter ranges. Within this setup, we obtain precise conditions for spin filtering by numerically deriving the spin polarized electric currents generated by electrons drawn from an unpolarized reservoir as a function of temperature and chemical potentials. The experimental conditions are derived using the Landauer-Buttiker transport formalism within a tight-binding model that incorporates realistic material parameters and the effects of voltage reservoirs. This description affords specific experimental parameters for building the filtering device regarding spatial dimensions, strength of  material parameters and of the externally applied fields and biases, in order to achieve optimal spin polarized yields.  

It is important to point out that the translation operator approach to this problem in reference \cite{Lopez,Hatano}  only provides a conceptual proof that in principle interferometry can yield spin selection. There, exact solutions can be found ignoring many practical drawbacks that should be evaluated for a successful proposal of a device, namely: The treatment presented here accounts for a finite input band of energies due to leads connected to a reservoir and to an output sink. It also accounts for the effect of temperature through the Fermi function at the reservoirs as long as the coherence length does not become shorter than the device size. Finally, finite voltage differences are contemplated so that an actual current can be driven through the device. The latter source of control of the input energies lends itself as an additional tuning parameter to optimize spin filtering, a tuning not accessible in previous more idealized proposals. 

We present two general device configurations: i) a symmetric arm configuration, that requires a weak magnetic field and renders an almost flat energy dependence of the polarization conditions and ii) an asymmetric interferometer, that requires no external magnetic field and only Rashba couplings, with a more complex energy dependence that demands specific voltage ranges to filter. This proposal shows the practical possibilities, within a 2D electron gas,  of the interferometric concept advanced theoretically by Hatano, Shirasaki and Nakamura\cite{Hatano}\cite{Ting}. 

\section{Methods}
We consider a quasi two dimensional electron gas consisting of non interacting electrons subject to both Rashba and Dresselhaus spin orbit interactions. In addition, an external magnetic flux $\Phi_B$, described by a vector potential ${\vec A}$, threads the device transversally.  Recent works have shown how to measure and control the Rashba parameter using gate voltages in two dimensional GaAs/AlGaAs electron gas\cite{Nitta2,Shapers,Studer,MillerGoldhaberGordon} and also in other heterojunctions, such as InAs/AlSb and HgTe\cite{HeidaSchultz}. In general the Rashba parameter $\alpha=b\langle E\rangle$, where $\langle E\rangle$ is the expectation value of the electric field at the 2DEG, and $b$ depends on the inverse of both the effective mass and the material gap\cite{Lommer}.  Measurements of SO parameters have been made by either Shubnikov-de Haas or weak localization/antilocalization effects. While intrinsic Dresselhaus parameters cannot be changed by a gate potential, they can be tuned by strain effects\cite{LiLi}.

One can address the two dimensional GaAs/AlGaAs electron gas by a single particle Hamiltonian including the previously described couplings by
\begin{equation}\label{Hamiltonian}
{\bf H}= \frac{{\vec {\bf\Pi}}^2}{2m^*} + {\bf  U} - \frac{\alpha}{\hbar} (\Pi_x\bsigma^y-\Pi_y \bsigma^x)- \frac{\beta}{\hbar}(\Pi_y\bsigma^y-\Pi_x\bsigma^x)+ \frac{\hbar \omega_B}{2}\bsigma^z, 
\end{equation}
where ${\vec {\bf\Pi}}=({\vec {p}}+e{\vec A})\nbOne$ is the kinetic momentum, 
${\nbOne}$ being the $2\times 2$
identity matrix (in spin space). Ordinary vectors are represented with an 
over-arrow while $2\times 2$ matrices are written in bold face.
The electron's charge and effective mass are denoted as $-e$ and $m^*$, 
${\bf U}=U\nbOne$ is a substrate lattice periodic potential, $\bsigma$ 
is a vector of Pauli matrices, and $\alpha$ and $\beta$ are material-dependent parameters characterizing the Rashba and Dresselhaus interactions, respectively\cite{Halperin}. 
The simple form for the Dresselhaus term is due to the assumption of strong confinement of the electron gas, so that $k_F << \pi/d$\cite{Halperin} where $k_F$ is the Fermi wavector of in plane electrons and $d$ in the confinement length scale. The condition essentially states that the direction perpendicular to the gas can be averaged, reducing cubic contributions to the dominant linear terms described here. The Dresselhaus term depends on the orientation of the crystal axes, our geometry insures that all interferometer arms are described by the same Hamiltonian.
The last term is the Zeeman energy, which we will ignore in the limit of small magnetic fields (a few flux quanta through a ${\rm 200}\times {\rm 200} \mu {\rm m}^2$ area) where it is much smaller than the spin orbit energy\cite{MillerGoldhaberGordon}\cite{Takayanagi}.
  
\begin{figure}
\begin{center}
\includegraphics[width=8 cm]{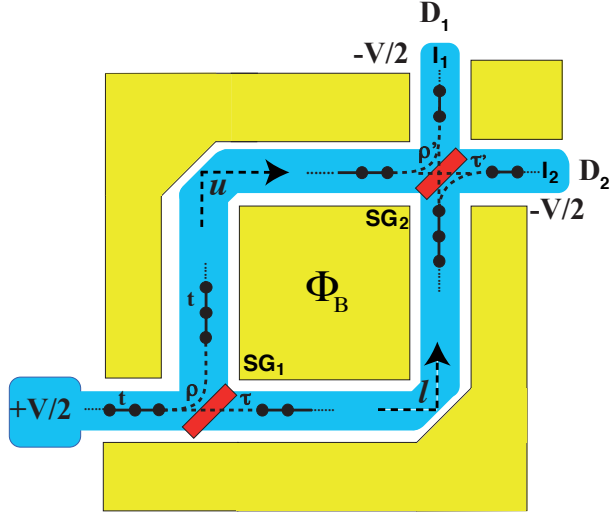}
\end{center}
\caption{Electronic Mach Zenhder interferometer setup on a 
SO active GaAs/AlGaAs 2D electron gas and the corresponding tight-binding model superposed. The device is biased on the left 
at voltage $+V/2$ and the current is collected on the right at the two 
reservoirs biased at $-V/2$. The electron beam splitters are implemented 
through two Schottky gates with reflection and transmission coefficients $\rho$ and $\tau$ separating electrons between upper $u$ and lower $l$ tight binding arms. There is a magnetic flux $\Phi_B$ perpendicular to the plane. The outgoing currents $I_1$ and $I_2$ are collected at $D_1$ and $D_2$.}
\label{fig1}
\end{figure}

For a GaAs heterostructure $\alpha\sim 3.9\times 10^{-12}{\rm eV~ m}$\cite{DattaDas}, $\beta\sim 2.4\times 10^{-12} {\rm eV~m}$ and $\hbar^2/m^* L\sim 1.7 \times 10^{-12}{\rm eV~ m}$, assuming electron ballistic propagation of length $\sim 1 \mu m$ and an effective mass of $m^*=0.067 m_0$, with $m_0$ the free electron mass. The proposed device configuration is depicted in Fig.\ref{fig1}, a Mach Zehnder Interferometer (MZI) implemented by voltage gating two interfering electron paths. GaAs heterostructures are known to have charge phase coherence lengths of 20$\mu$m at 15mK while spin coherence lengths are larger: $\sim$ 100$\mu$m at 1.6K\cite{Ionicioiu}. Thus, at 15mK  both coherence lengths will be larger than a device of 20$\mu$m size.

We are interested in determining the polarized currents $I_{D_i}$ at outgoing channels $D_i$, with $i=1,2$ and to find the conditions for spin filtering\cite{Lopez} at either output channel. We define the spin filter as one acting on any entering polarization and returning a polarized state along a definite axis. This approach will serve to build two polarized spin currents of opposite polarization. The relevant processes within the interferometer are described as follows (see figure \ref{fig1}): Single electrons are assumed to be extracted from the left electron reservoir at voltage +V/2. The electrons then pass through the first beam splitter, implemented by a Schottky gate\cite{OliverYamamoto} labeled (${\rm SG_1}$)\cite{Yamamoto} resulting in two intermediate output beams. These two beams interfere at the second Schottky gate (${\rm SG_2}$), from which two final beams emerge at the output channels.
 
As we consider both Rashba and Dresselhaus interactions, we need to derive general reflection conditions at the Schottky gates. In reference \cite{Yamamoto}, this was done for the Rashba interaction, assuming that small enough spin-orbit strength would yield only a small divergence of the reflected spin states. In this paper we will take the case of $\pi/4$ reflections, that can be controlled experimentally, and lead to simple, spin-orbit independent reflection and transmission matrix elements\cite{Lopez}. If it were not the case and large deviations from such conditions arose the beam splitter would behave differently as the spin-orbit parameters changed and would complicate predicting an appropriate operating regime.

A tight binding model will allow us to quantitatively parametrize for the two dimensional gas and spin-orbit intensities appropriate to a GaAs/AlGaAs 
junction. 
The tight-binding Hamiltonian for the interferometer SO active region is
\begin{widetext}
\begin{eqnarray}
{H}^{\rm TB}&=&\sum_{{\vec r}}(c^\dag_{\vec r\uparrow}\ 
c^\dag_{\vec r\downarrow})\begin{pmatrix}\epsilon_{\vec r} &0\\
0&\epsilon_{\vec r}\end{pmatrix}
\begin{pmatrix}c^\dag_{\vec r\uparrow}\\
c^\dag_{\vec r\downarrow}\end{pmatrix}+
\sum_{{\vec r},\vec{r'}}(c^\dag_{\vec r\uparrow}\ 
c^\dag_{\vec r\downarrow})\begin{pmatrix}
t_{\vec r\vec{r'}}^{\uparrow\uparrow} &t_{\vec r\vec{r'}}^{\uparrow\downarrow}\\
t_{\vec r\vec{r'}}^{\downarrow\uparrow}&t_{\vec r\vec{r'}}^{\downarrow\downarrow}\end{pmatrix}
\begin{pmatrix}c^\dag_{\vec {r'}\uparrow}\\
c^\dag_{\vec {r'}\downarrow}\end{pmatrix}\nonumber\\
&=&\sum%^{(\uparrow,\downarrow)}
_{{\vec r}\lambda}\epsilon_{\vec r}{c}^{\dagger}_{{\vec r}\lambda}
{c}_{\vec r\lambda}
+\sum%^{(\uparrow ,\downarrow)}
_{{\vec r}\lambda,\vec {r'}\lambda^\prime}
{t}_{{\vec r}\vec {r'}}^{\lambda\lambda^\prime}{c}^{\dag}_{{\vec r}\lambda}{c}_{\vec {r'} \lambda^\prime},\label{eqTB}
\end{eqnarray}
\end{widetext}
where $\epsilon_{{\vec r}}$ are the site energies, ${c}^{\dagger}_{{\vec r}\lambda}$ creates electrons at site ${\vec r}$ with spin $\lambda=\uparrow,\downarrow$ and ${t}_{{\vec r}\vec {r'}}^{\lambda\lambda^\prime}$ is the transfer integral between sites/spins ${\vec r},\lambda$ and $\vec {r'},\lambda'$. The sum over the sites $\vec r$ draws the legs of the interferometer, like in 
Fig.~\ref{fig1}.  The states are denoted as  $|\dots,n_{\vec r\lambda},\dots\rangle$ with
$n_{{\vec r}\lambda}=0,1$ depending on whether or not  an electron with spin $\lambda$
occupies site $\vec r$. The link between equations (\ref{Hamiltonian}) and (\ref{eqTB}) is established
through the definition of the spinor components $\psi^\lambda (\vec r)=\langle\vec r\lambda|\psi\rangle$.

The transfer integral between sites $\vec r$ and $\vec {r'}$ 
including both field and SO effects measures the phase aquired by the electrons
when they are transported along the interferometer legs. 
It can be written in the ``position-spin'' basis $\{|\vec {r}\lambda\rangle\}$ 
as
\begin{equation}
{t}_{\vec r\vec {r'}}^{\lambda\lambda'}=\langle\vec r\lambda|
t~\!\exp\left[\left(\frac{2 \pi i}{\phi_0}
\right)\vec{\pmb{\cal{A}}}
((\vec{r}+\vec {r'})/2)\cdot (\vec{r}-\vec {r'})\right]|
\vec {r'}\lambda'\rangle,\label{eqt}
\end{equation}
where $t$ is the coupling between sites and it is set to 0.156 eV (see Fig.\ref{fig1}), 
in accordance to the Fermi velocity in GaAs, $\phi_0=h/e$ 
is the flux quantum and  
\begin{eqnarray}
\vec{\pmb{{\mathcal{A}}}}(x,y)
&=&\left[-\frac{1}{2}B_z y\nbOne+\frac{m^*}{e} (\beta \bsigma^x-\alpha \bsigma^y)
\right]\vec u_x\\
&+& \left[\frac{1}{2}B_z x\nbOne 
+\frac{m^*}{e}(\alpha \bsigma^x-\beta \bsigma^y)\right]\vec u_y
\label{eqA}
\end{eqnarray}
accounts for the spin spatial-precession and phase changes due to the SO coupling and the magnetic
field, that operates in between tight binding sites e.g. at the middle of the link which connects sites $\vec r$ and $\vec {r'}$.
We have chosen the simple staggered gauge for the magnetic field vector potential. 
Equations~(\ref{eqt}) and (\ref{eqA}) are easily derived 
when the SO interaction is described in terms of a minimal 
coupling~\cite{Jin,Leurs,Medina,Mineev,Frohlich,Tokatly}. 

For the current incoming from the left electron reservoir and outgoing to the detectors, we have transmission and reflection coefficients 
$\tau,\rho$ and $\tau',\rho'$ respectively (see Fig.\ref{fig1}), that can be gate controlled. The gate is spin inactive as it constitutes a scalar potential barrier, for both the input and output channels, so that in the notation of Eq.\ref{eqt}, $t_{in,r_{l0}}^{\lambda,\lambda}=\tau$, $t_{in,{\vec r}_{u0}}^{\lambda,\lambda}=\rho$, $t_{{\vec r}_{uN},D_1}^{\lambda,\lambda}=\rho'$, $t_{{\vec r}_{uN},D_2}^{\lambda,\lambda}=\tau'$, $t_{{\vec r}_{lN},D_1}^{\lambda,\lambda}=\tau'$, $t_{{\vec r}_{lN},D_2}^{\lambda,\lambda}=\rho'$ where ${\vec r}_{u0}$ and ${\vec r}_{l0}$ are the first tight-binding sites of the arms and ${\vec r}_{uN}$ and ${\vec r}_{lN}$ are the $N$th sites of the corresponding arms. We choose throughout $\rho=\rho'=i/\sqrt{2}$ and $\tau=\tau'=1/\sqrt{2}$.

The coupling to the metallic reservoirs is described by a self-energy~\cite{Pastawski}. The self energy is derived by solving
the Dyson equation for a chain of sites indexed by $n$ where $\Sigma_n= t\frac{1}{E-\Sigma_{n+1}}t$ and $\Sigma_n=\Sigma_{n+1}$, for site energies chosen to be zero, for an  infinite chain. The resulting quadratic equation can be solved and has real and imaginary parts so that
\begin{equation}
{  \Sigma}_{\lambda}^{(lead)}(E)=\frac{E-i\sqrt{4t^2-E^2}}{2}
{c}^\dag_{{\rm lead}\lambda}{c}^{\phantom{\dag}}_{{\rm lead}\lambda}, 
\label{eq:autoenergia}
\end{equation}
where we have introduced creation and annihilation electron states at each lead of the system i.e. $in$, $D_1$, and $D_2$ for the incoming lead and the two drains, where a metallic lead couples the system to the reservoirs. Three such couplings
will be considered below, at the incoming beam and at the two output leads connecting to detectors.
The transmission between a spin state $\lambda$ at the input and spin state $\tau$ at the output $D_i$, at energy $E$ is given by\cite{Datta}
\begin{eqnarray}
T_{D_j}^{\tau\lambda}&=&{\rm Tr} \left [{ \Gamma}_{\tau}^{D_j} 
{  G}(E)~ { \Gamma}_{\lambda}^{in}~{  G}(E)^{\dagger}\right ],
\label{Transmission}
 \end{eqnarray}
 where the broadening, due to the coupling to the electron reservoirs is given by
${ \Gamma}_{\tau}^{D_j}= i\left [{ \Sigma}_{\tau}^{D_j} -(
{ \Sigma}_{\tau}^{D_j})^{\dagger}\right ]$ and the Green's function is computed as
 \begin{equation}
{  G}(E)=\frac{1}{E-{H}^{TB}-({ \Sigma}^{in}+{ \Sigma}^{D_1}
+{ \Sigma}^{D_2})},
\label{Greens}
 \end{equation}
 where the self energies are diagonal matrices that include all spin orientations at the corresponding leads~\cite{Datta}. The current at each output lead $D_j$ with spin component $\tau$ is computed using
\begin{equation}
I^{\tau}_{D_j}=\frac{e^2}{h}\sum_{\lambda}\int_{band} T^{\tau\lambda}_{D_j}(E)\left ( f^{in}(E)-f^{D_j}(E)\right ) dE.
\label{I-Vequation}
\end{equation}
The Fermi occupation at temperature $T$ is given by $f^i(E)=1/(\exp(\frac{E-E^i_F}{k_BT})+1)$, where $E^i_F=\pm V/2$.  

The numerical calculation would proceed in the following way: The incoming and outgoing one dimensional leads, are made up from site energies set to zero as a reference energy and nearest neighbor couplings $t$ set to 0.156 eV. The self energy of such leads can be computed 
exactly\cite{Pastawski} leading to an energy contribution at the input and output sites of the interferometer. The Hamiltonian of the interferometer is then computed according to Eq.\ref{eqTB} and the hopping matrix elements in Eq.\ref{eqt}. The spin orbit parameters $\alpha$ and $\beta$ are only different from zero within the interferometer. One can then combine the computed matrices for the Hamiltonian and the self energies, into the Green's function defined by Eq.\ref{Greens} which involves a matrix inversion and contains all the information on the external contacts of the interferometer. Such an expression is also a matrix and is used to compute the transmission between specific spin components through the interferometer by way of Eq.\ref{Transmission}. The transmission is a number once the spin components have been fixed, and it must be computed for each of the energies within the incoming band. From this function, the spin polarized currents can be obtained.

As a measure of the performance of the device we use the transmission 
asymmetry at a particular output channel, which is defined as
\begin{equation}
A_{D_j}(\phi,\alpha,\beta)=T_{D_j}^{\uparrow\uparrow}+T_{D_j}^{\downarrow\uparrow}-(T_{D_j}^{\downarrow\downarrow}+T_{D_j}^{\uparrow\downarrow}),
\label{equationasymmetry}
\end{equation}
considering that the states at the input lead are thermal mixed states at temperature $T$. In this case, 
the amplitudes do not interfere and the calculation requires summing probabilities.  
This measure is used due to the fact that, in contrast to the translation operator approach of previous proposals, we cannot obtain exact conditions for spin filtering when reservoirs are considered, although the arms of the device are one dimensional. Therefore, the conditions for operation must be sought numerically, with an adequate measure for performance.

\section{Results}
\begin{figure*}
\includegraphics[width=12 cm]{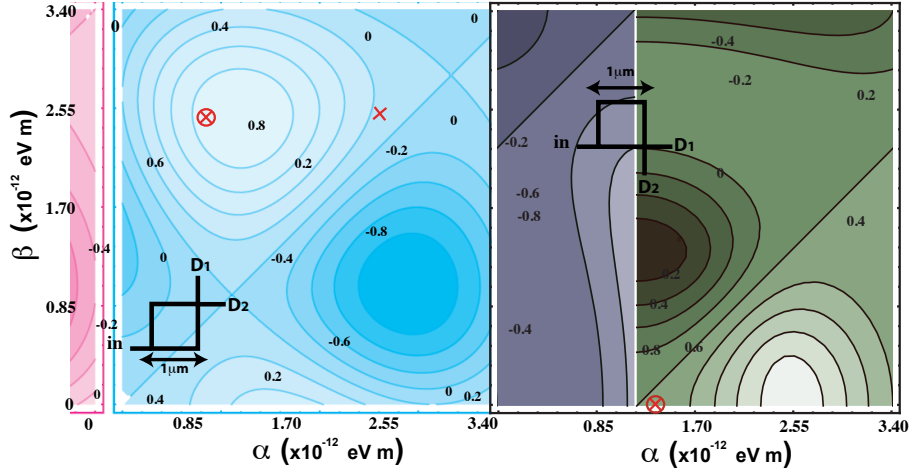}
\caption{Transmission asymmetry as a function of the Rashba and Dresselhaus spin-orbit interaction. The clearer shades represent positive asymmetry indicating up spin preference according to Eq.\ref{equationasymmetry}, the values on the lines represent the magnitude of the asymmetry. Left panel: The symmetric interferometer (see inset), where the point of operation is indicated by a $\bigotimes$ at values encountered in GaAs/AlGaAs heterostructures\cite{Takayanagi} while {\LARGE \bf $\times$} is a representative point for small asymmetry. The magnetic flux at the operation point is $\Phi=0.248 (h/e)$ a field of 10 Gauss through an interferometer of 1$\mu$m$^2$. Right panel: The asymmetric interferometer (see inset) where the point of operation is indicated by a $\bigotimes$. There is no applied magnetic flux at the operation point and the Dresselhaus parameter can be close to zero for an interferometer of 1$\mu$m$^2$.}
\label{fig2}
\end{figure*}
Figure \ref{fig2} shows the spin filtering asymmetry at output lead $D_1$ as a function of the Rashba and Dresselhaus values for the material. The magnetic field has been chosen so that the spin asymmetry is maximized. The potential points for operation are identified in the figure within less than one flux quantum through the device. Such operation point comprises an asymmetry of more than 99\% so the filter is highly efficient. Furthermore, the sensibility of the filtering asymmetry to changes in the parameters is weak, i.e. a ~15\% change in Rashba coupling only changes the asymmetry in 10\% around the operating point. Thus, tuning around the operation point is by no means beyond experimental accuracy.
\begin{figure}
\begin{center}
\includegraphics[width=8 cm]{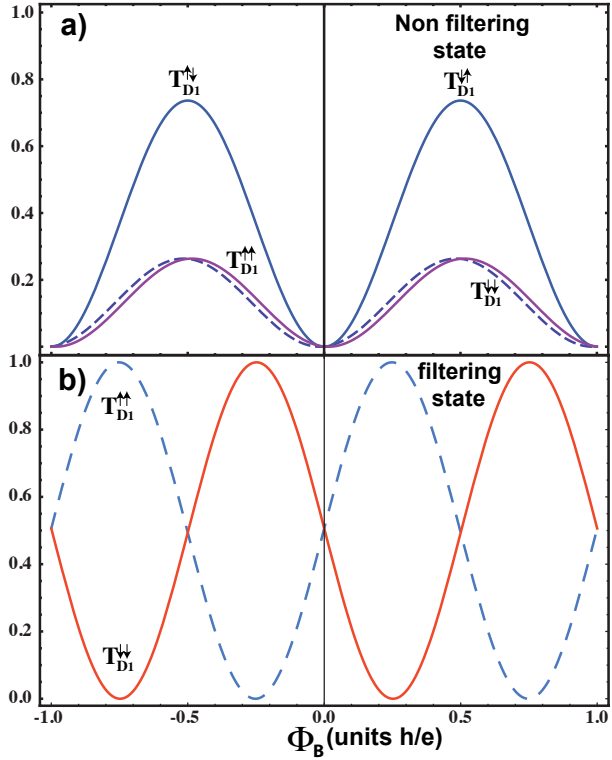}
\end{center}
\caption{a) Transmission probabilities as a function of magnetic flux for the point {\LARGE $\times$} in Fig.\ref{fig2} left panel, off the filtering operating point. b) Transmission probabilities $T_{D_1}^{\uparrow\uparrow}$ and $T_{D_1}^{\downarrow\downarrow}$ at lead $D_1$ at the operation point $\bigotimes$ (see Fig.\ref{fig2} left panel), as a function of the magnetic flux. The transmission coefficients  $T_{D_1}^{\uparrow\downarrow}$ and $T_{D_1}^{\downarrow\uparrow}$ are negligible at the operation point. The behavior of the lead $D_2$ is exactly opposite to that of $D_1$ allowing for a separation of opposite components in a single device.}
\label{fig3}
\end{figure}

Figure \ref{fig3} shows the transmission at selected points in the $\alpha,\beta$ parameter space (see Fig.\ref{fig2}) as a function of the magnetic flux through the device from the input to the output channel $D_1$. The top panel depicts an $\alpha,\beta$ choice with negligible filtering properties, where all transmission probabilities are appreciable so that only a weak asymmetry results. 
The bottom panel shows the transmission probabilities at the filtering operating point. The figure shows that one can choose the magnetic flux so as to obtain optimal filtering for either spin component (there are two optimal operating points in the range chosen for the SO parameters). The other components for the specific operation point are negligible for all fields shown. The signal at output channel $D_2$ will show the opposite behavior as depicted by the dashed blue lines in the bottom panel $(b)$ of Fig.\ref{fig3} (exchanges $\uparrow\uparrow$ for $\downarrow\downarrow$), so we can simultaneously and separately draw both spin orientations. We note that there are no optimal operating points for either $\alpha~ {\rm or}~ \beta=0$, for which only a small asymmetry results.

An important feature of the proposed device is that, in the symmetric arm configuration, the filtering properties are robustly energy independent in the tight-binding model i.e. the properties remain in the full band width of the incoming channel\cite{Hatano}. Small interferometer arm asymmetries are nevertheless inevitable in a real setup, so
as we will see below, an energy dependence is expected. 
Due to the insensitivity of the filtering to the input energy, the polarized current versus applied voltage is linear (see Eq. \ref{I-Vequation}) whenever $k_B T\ll eV$ 
i.e. the thermal energy is small compared to the potential difference applied to the device. 
\begin{figure}
\begin{center}
\includegraphics[width=8 cm]{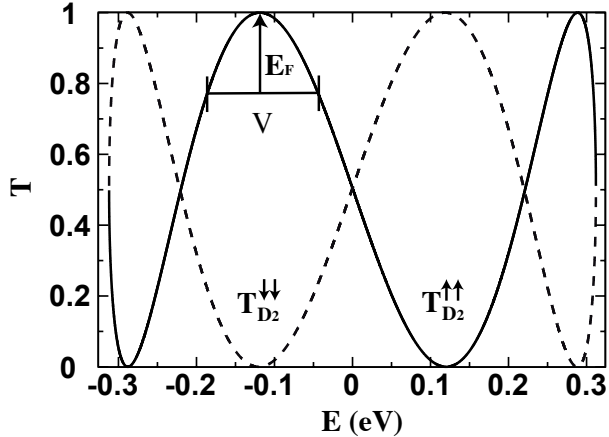}
\end{center}
\caption{Transmission probabilities $T_{D_2}^{\uparrow\uparrow}$ and $T_{D_2}^{\downarrow\downarrow}$ at lead $D_2$ as a function of energy at the operation point ${\bigotimes}$ in Fig.\ref{fig2} right panel. The transmission coefficients  $T_{D_2}^{\uparrow\downarrow}$ and $T_{D_2}^{\downarrow\uparrow}$ are negligible at the operation point. The input gate potential (Fermi energy) is centered at the peak of transmission for the {\it up spin} in $D_2$, then the potential difference between the input and lead is increased as shown.}
\label{fig4}
\end{figure}

The second configuration we analyze consists of an interferometer with a large asymmetry in arm lengths. In this configuration one can filter in the absence of a magnetic field using the difference in translational phase. As we have fixed the length difference, one does not achieve as high a filtering capability as in the symmetric arm configuration. This setup also compromises the energy dependence i.e. it is only for certain ranges of energy, within the input band, that the interferometer will work. Nevertheless, arm lengths could also be tuned to yield a higher asymmetry. Figure \ref{fig2} right panel, shows the transmission probability asymmetry for the second interferometer. Again, one can tune the SO parameters in the same range as in the symmetric arm setup. Nevertheless, for the particular configuration chosen (a 3$\mu$m and a 1$\mu$m arm) the optimal point of operation occurs at zero Dresselhaus coupling. The latter coupling could in principle be achieved by means of strain\cite{LiLi}.  

\begin{figure}
\begin{center}
\includegraphics[width=8 cm]{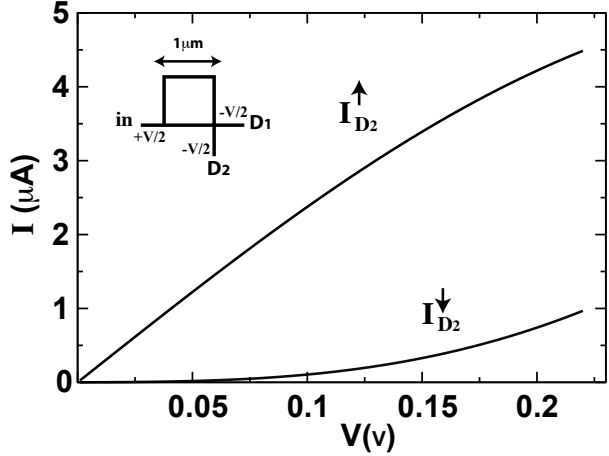}
\end{center}
\caption{Polarized current components at lead $D_2$ as a function of the potential difference as indicated in Fig.\ref{fig4}. The applied magnetic flux is zero. Above 0.1 V there is an increasing contribution from the opposite spin orientation that degrades polarization. The inset shows the proposed configuration that differs from that of Fig.\ref{fig1} by the position of the output leads.}
\label{fig5}
\end{figure}

In Figure \ref{fig4} we show the transmission probability as a function of the energy for the spin flipping processes ($\uparrow\rightarrow\downarrow$ and $\downarrow\rightarrow\uparrow$), the other components being negligible. The transmission is no longer optimal in the whole energy range but peaks in certain energy intervals. If one tunes the input Fermi energy and then samples potential difference in the vicinity of the optimal transmission, for a particular output lead, one can obtain a substantial polarized current within a certain bias range. Figure \ref{fig5} shows the polarized current components for the situation depicted in Fig.\ref{fig4}. The figure shows that one can achieve at least four times as much {\it spin up} polarized current as {\it spin down} current contribution within a 0.2 volt bias around an optimal transmission probability.
\section{Summary}
We have proposed a practical spin filtering device based on a Mach-Zehnder type spin interferometer built into a 2D GaAs/AlGaAs electron gas. Spin filtering operation is achieved by separating spin up and spin down components of an incoming thermal mixed state extracted from a biased electrode. Spin filtering is possible by tuning Rashba and/or Dresselhaus spin-orbit contributions by way of an applied gate voltage or strain.  The Landauer-Buttiker formulation, properly accounting for coupling to reservoirs at finite temperatures and voltage bias, is used to compute transmissions and polarized currents and demonstrate the robustness of the interference concept for current polarization. Two device modes are discussed that operate at complementary experimental conditions: i) A symmetric device that requires a weak magnetic field and tuning both Dresselhaus and Rashba interactions. Such configuration displays a flat incoming electron energy dependent filtering, and ii) an asymmetric device that only requires tuning of Rashba couplings and operates at specific voltage biases. Such energy dependence makes the latter device a simpler setup but more prone to voltage and temperature sensitivities.  Both configurations display the drawing of opposite spin polarized currents at the output leads at the optimal operation point.

While the gate voltage control is well a established mechanism, strain has recently been tested for spin control by building mismatched layers in AlGaAs\cite{Jain} generating spin splittings of the BIA type of around 0.007 meV, and 0.025 meV for the bulk Dresselhaus interaction. Strain can account for a 20\% change in the BIA coupling in mismatched AlGaAs\cite{Bernevig3}. This is within the requirements of the proposed device. An additional tunable parameter is a weak perpendicular external magnetic field that induces effects similar to tuning the interferometer arm lengths. The interferometric setup, proposed here, avoids using ferromagnetic material or strong magnetic fields (Zeeman term or spin torque mediated) to polarize electron spins. The device also operates at broad energy ranges for the incoming electron band, a fact which averts the need for difficult fine tuning of the required parameters. 

\acknowledgments
This work was supported by CNRS-Fonacit grant PI-2008000272.


\section*{References}

\begin{thebibliography}{10}
\bibitem{Rashba}  E. I. Rashba, {\it Sov. Phys. Solid State} {\bf 2}, 1109 (1960).
\bibitem{Dresselhaus} G. Dresselhaus,  {\it Phys. Rev.} {\bf 100}, 580 (1955).
\bibitem{winkler} R. Winkler,  {\it Spin-Orbit Coupling Effects in Two Dimensional Electron and Hole Systems} (Springer) (2003).
\bibitem{Nitta} J. Nitta  and T. Koga,  {\it J. Supercond.} {\bf 16}, 689 (2003).
\bibitem{Ionicioiu} R. Ionicioiu  and I. D'Amico,   {\it Phys. Rev. B} {\bf 67}, 041307(R) (2003).
\bibitem{Hatano} N. Hatano,  R. Shirasaki  and H. Nakamura,  {\it Phys. Rev. A} {\bf 75}, 032107 (2007).
\bibitem{SHChen}  S -H Chen and C -R Chang,  {\it Phys. Rev. B.} {\bf 77}, 045324 (2008).
\bibitem{SarmaReview} I. Zutic, J. Fabian and S. Das Sarma,  {\it Rev. Mod. Phys.} {\bf 76}, 323 (2004).
\bibitem{BalseiroUsaj} G. Usaj and C. A. Balseiro,  {\it Europhys. Lett.} {\bf 72}, 631 (2005).
\bibitem{Koga} T. Koga, J. Nitta and M. van Veenhuizen,  {\it Phys. Rev. B} {\bf 70}, 161302(R) (2004).
\bibitem{Bernevig2} B. A. Bernevig, J.  Orenstein and  S. C. Zhang,  {\it Phys. Rev. Lett.} {\bf 97}, 236601 (2006).
\bibitem{ZulickeAlone} U. Zulicke,  {\it Appl. Phys. Lett.} {\bf 85}, 2616 (2004).
\bibitem{SignalZulicke} A. I. Signal and U. Zulicke ,  {\it Appl. Phys. Lett.} {\bf 87}, 102102 (2005).
\bibitem{Lopez} A. L\'opez, E. Medina, N.  Bol\'\i var and B. Berche, {\it J. Phys.: Condens. Matter} {\bf 22},  115303 (2010).
\bibitem{Berche1} B. Berche, N. Bol\'ivar, A. L\'opez and E. Medina,
 {\it Cond. Matt. Phys.} {\bf 12}, 707  (2009).
\bibitem{Ting}  D Z -Y Ting and X. Cartoixa,  {\it Phys. Rev. B.} {\bf 68}, 235320 (2003).
\bibitem{Nitta2} J. Nitta, T. Akazaki, H. Takayanagi and T. Enoki, {\it Phys. Rev. Lett.}{\bf 78}, 1335 (1997).
\bibitem{Shapers} T. Shapers et al, {\it J. Applied Phys.} {\bf 83}, 4324 (1998).
\bibitem{Studer} M. Studer, G. Salis, K. Ensslin, D. C. Driscoll and A. C. Gossard,  {\it Phys. Rev. Lett.} {\bf 103}, 027201 (2009).
\bibitem{MillerGoldhaberGordon} J. B. Miller, D. M. Zumbuhl, C. M. Marcus, Y. B. Lyanda-Geller, D. Goldhaber-Gordon, K. Campman and A. C. Gossard,  {\it Phys. Rev. Lett.} {\bf 90}, 076807 (2003).
\bibitem{HeidaSchultz} J. P. Heida, B. J.  vanWees, J. J. Kuipers, T. M. Klapwijk, and G. Borghs, {\it Phys. Rev. B}{\bf 57}, 11911 (1998); M. Shultz et al, {\it Semiconduc. Sci. Technol}, {\bf 11}, 1168 (1996).
\bibitem{Lommer} G. Lommer, F. Malcher, and U. Rossler, {\it Phys. Rev. Lett.}, {\bf 60}, 728 (1988).
\bibitem{LiLi} Y. Li, and Y-Q Li , {\it Eur. Phys. J. B}, {\bf 63}, 493 (2008).
\bibitem{Halperin} H. A. Engel, E. I. Rashba and B. I. Halperin, {\it Theory of Spin Hall Effects in Semiconductors}, in Handbook of Magnetism and Advanced Magnetic Materials, H. Kronm\"uller and S. Parkin (eds.). (John Wiley \& Sons Ltd, Chichester, UK) (2007).
\bibitem{Takayanagi} J. Nitta, F. E. Meijer and H. Takayanagi,  {\it Appl. Phys. Lett.} {\bf 75}, 695 (1999).
\bibitem{DattaDas}  S. Datta and B. Das, {\it Appl. Phys. Lett.} {\bf 56}, 665
(1990).
\bibitem{OliverYamamoto} W. D. Oliver, J. Kim, R. C. Liu and Y. Yamamoto, {\it Science} {\bf 284}, 299 (1999).
\bibitem{Yamamoto} G. Feve, W. D. Oliver, M. Aranzana, and  Y. Yamamoto,  {\it Phys. Rev. B} {\bf 66}, 155328 (2002).
\bibitem{Jin} P. Q. Jin, Y. Q. Li and F. C. Zhang,  {\it J. Phys. A: Math. Gen.} {\bf 39}, 7115 (2006).
\bibitem{Leurs} B. W. A. Leurs, Z.  Nazario, D. I.  Santiago and J. Zaanen,  {\it Ann. Phys.} {\bf 323}, 907 (2008).
\bibitem{Medina} E. Medina, A. L\'opez and B. Berche, {\it Europhys. Lett.} {\bf 83}, 47005 (2008).
\bibitem{Mineev} V. P. Mineev and G. E. Volovik,  {\it J. Low Temp. Phys.} {\bf 89}, 823 (1992).
\bibitem{Frohlich} J. Fr$\ddot{\rm o}$hlich and U. M. Studer,  {\it Rev. Mod. Phys.} {\bf 65}, 733 (1993).
\bibitem{Tokatly} I. V. Tokatly,  {\it Phys. Rev. Lett.} {\bf 101}, 106601 (2008)
\bibitem{Pastawski} H. M. Pastawski and E. Medina, {\it Revista Mexicana de F\'isica} {\bf 47}, Suppl. 1, 1 (2001). (http://rmf.smf.mx/pdf/rmf-s/47/1/47\_1\_0001.pdf)
%\bibitem{Sasada05}Sasada K. and Hatano N., 
%{\it Physica E} (Amsterdam) {\bf 29}, 609 (2005).
\bibitem{Datta} S. Datta, {\it Electronic transport in mesoscopic systems},
Cambridge University Press, Cambridge 1995.
%\vskip10mm
%\bibitem{Goldhaber} A. S. Goldhaber, {\it Phys. Rev. Lett.} {\bf 62}, 482 (1989).
%\bibitem{ryder} L. H. Ryder,  {\it Quantum Field Theory}, (Cambridge University Press) (1985).
%\bibitem{Rebei} A. Rebei and O. Heinonen,  {\it Phys. Rev. B} {\bf 73}, 153306 (2006).
%\bibitem{AharonovCasher} Y. Aharonov and A. Casher, {\it Phys. Rev. Lett.}
%	{\bf 53}, {319} (1984).
%\bibitem{Comment} A. L\'opez, E. Medina, N. Bol\'ivar and B. Berche, {\it Preprint} arXiv:cond-mat/0902.4635.
%\bibitem{Peskin} M. E. Peskin and D. V. Schroeder,  {\it Quantum Field Theory} (Westview) (1995).
%\bibitem{Thornton} T. J. Thornton, M. Pepper, H. Ahmed, D. Andrews and G. J. Davies, {\it Phys. Rev. Lett.} {\bf 56}, 1198 (1986).
\bibitem{Jain} S. C. Jain, H. Willander and H. Maes, {Semicond. Sci. Technol.}, {\bf 11}, 641 (1996).
\bibitem{Bernevig3} B. A. Bernevig and S. C. Zhang, {\it Physical Review B}, {\bf 72}, 115204 (2005).



\end{thebibliography}
\end{document}